\documentclass{aa}
\usepackage{graphicx,epsfig}
\usepackage{amssymb,amsmath}
\usepackage{xspace}
\usepackage{natbib}
\usepackage{ulem}
\bibpunct{(}{)}{;}{a}{}{,}
\newcommand{\etal}{{et al.\ }}

\begin{document}

%
\title{Spectral and Photometric Evolution of Young Stellar Populations: the Impact of Gaseous Emission at Various Metallicities\thanks{Data files are provided in the electronic version/at CDS, and at http://www.uni-sw.gwdg.de/$\sim$galev/panders/} } 

\author{Peter Anders, Uta Fritze -- v. Alvensleben}
\institute{Universit\"ats-Sternwarte, Geismarlandstr.~11, 37083 G\"ottingen, Germany; Email: panders@uni-sw.gwdg.de, ufritze@uni-sw.gwdg.de}
\authorrunning{P. Anders \& U. Fritze -- v. Alvensleben}
\titlerunning{Young Stellar Populations}
\date{Received xxx  / Accepted xxx}

\abstract{We include gaseous continuum and line emission into our {\bf {\sc galev}} models for the spectral and photometric evolution of {\bf S}imple {\bf S}tellar {\bf P}opulations ({\bf SSP}s) for various metallicities in the range 0.02 $\leq Z/Z_{\odot} \leq$ 2.5. This allows to extend them to significantly younger ages than before. They now cover the age range from  4 Myr all through 14 Gyr. We point out the very important contributions of gaseous emission to broad band fluxes and their strong metallicity dependence during very early evolutionary stages of star clusters, galaxies or subgalactic fragments with vigorous ongoing star formation. Emission-line contributions are commonly seen in these actively star-forming regions. Models without gaseous emission cannot explain their observed colors at all, or lead to wrong age estimates. We use up-to-date Lyman continuum {\bf(= Lyc)} emission rates and decided to use recent empirical determinations of emission line ratios relative to ${\rm H_{\beta}}$ for subsolar metallicities. We justify this approach for all situations where no or not enough spectral information is available to determine all the parameters required by photoionization models. The effects of gaseous line and continuum emission on broad band fluxes are shown for different metallicities and as a function of age. In addition to the many filter systems already included in our earlier models, we here also include the {\sl HST} NICMOS and Advanced Camera for Surveys (= ACS) filter systems.
}

\maketitle
%

\section{Introduction}
In Schulz \etal (2002) we presented spectral evolutionary synthesis models for SSPs of a wide range of metallicities 0.02 $\leq Z/Z_{\odot} \leq$ 2.5. Our models are based on isochrones from the Padova group that include the TP-AGB phase and on the spectral model atmosphere libraries from Lejeune \etal  (1997, 1998), rectified to yield agreement with observed colors from $U$ through $K$ of stars with effective temperatures in the range 2\,000 K through 50\,000 K. Stars with higher temperatures are treated as black body radiators. These models give the time evolution of spectra as well as of luminosities and colors in a large set of filter systems. As they did not include gaseous emission important during the lifetime of massive ionising stars, they cover a range of ages from 140 Myr through 14 Gyr.

SSP model results are useful not only for analyses of star clusters, the genuine SSPs, but are also readily superposed to describe the evolution of composite stellar populations like galaxies with star formation histories extended in time and with various chemical enrichment histories. They can also directly be combined with cosmological structure formation scenarios that include a star formation criterium.

Systems with active ongoing star formation like very young star clusters, actively star-forming or star-bursting galaxies in the local as well as in the high-redshift Universe, protogalaxies and subgalactic fragments in particular, require the inclusion of gaseous emission. This is not only important for the characteristic emission lines in the spectra but gaseous emission, both in terms of lines and continuum, give important contributions to broad band luminosities and colors (cf. Kr\"uger \etal 1995, Zackrisson \etal 2001).

The wealth of {\sl HST} ACS data currently becoming available -- on many very young star cluster systems in particular -- prompted us to not only include the gaseous emission in an updated way into our spectral evolutionary synthesis models for SSPs of various metallicities but also to provide versatile tables for the luminosity and color evolution of these SSPs in ACS filter bands.

Results of these models have already been proven very useful in the interpretation of the ACS Early Release Observations of the very young star cluster systems in the Mice and Tadpole galaxies (de Grijs \etal 2002).

\section{Input Physics}
\subsection{SSP Models}
Except for the gaseous emission that was not yet included in our SSP models before, we use the same input physics as presented in detail in Schulz \etal (2002). This includes isochrones from the Padova group containing the TP-AGB phase and model atmosphere spectra from Lejeune \etal (1997, 1998), extending from 90 {\AA} through 160 $\mu$m, for five different metallicities $Z~{\rm =0.0004,~0.004,~0.008,~0.02=}~Z{\rm _{\odot} ~and ~0.05}$ or ${\rm [Fe/H] = -1.7,~-0.7,~-0.4,~0 ~and ~+0.4}$. Inclusion of the TP-AGB phase has been shown to be very important for colors like $V-I$ and $V-K$. As we have shown in Schulz \etal (2002), age-dating of star clusters on the basis of their $V-I$ colors, as often done for young cluster systems in interacting galaxies or merger remnants, can go wrong by a factor 2 (at fixed metallicity and extinction) if the TP-AGB phase is neglected in the models. 

We present our results for a Salpeter and a Scalo IMF as in Schulz \etal (2002).

\subsection{Gaseous Emission}
\label{sec_input_gas}
Gaseous emission is primarily related to very hot stars, i.e. to massive stars in early evolutionary phases of an SSP. However, hot white dwarfs (WDs) can also contribute in later stages (but see next subsection). Gaseous emission depends on metallicity in a two-fold way. First of all, stars get brighter and hotter on average in stellar populations at lower metallicity. The lifetimes of low mass stars get shorter, those of high mass stars get longer at lower metallicities as compared to solar. This affects the output rates in terms of hydrogen ionising photons, $N_{\rm Lyc}$, of a stellar population as well as its absorption line spectrum and, hence both the strengths of all gaseous emission lines in a uniform way and the gas continuum emission. Second, the chemical composition and the physical properties of the gas exposed to the same ionising radiation field determine the relative strengths of different emission lines. 

The physical properties and the chemical composition of the gas ionised by a bunch of stars of given metallicity are not known {\sl a priori}. They determine, however, the flux ratios of non-hydrogen element lines relative to ${\rm H_{\beta}}$. We therefore chose to assume that the gas has the same solar-scaled abundances as our single burst single metallicity stellar population and to use observationally determined emission line ratios for all non-hydrogen lines. In this respect our approach is similar to that of P\'erez-Gonz\'alez \etal (2002) who also use observed line ratios for their sample of strong ${\rm H_\alpha}$ emitting UCM survey galaxies. It differs from those of Charlot \& Longhetti (2001), Moy \etal (2001), or Zackrisson \etal (2001), who assume electron densities and temperatures for the gas at low metallicities and couple a photoionisation code to their evolutionary synthesis model. 

On the basis of effective temperature and bolometric luminosity (and hence radius and surface gravity) of every star in a given isochrone, its flux of hydrogen ionising photons ($N_{\rm Lyc}$) is calculated from up-to-date non-LTE expanding model atmospheres that take into account line-blanketing as well as stellar winds and recent temperature and gravity calibrations (Schaerer \& de Koter 1997, Vacca \etal 1996, Smith \etal 2002). Summing up the $N_{\rm Lyc}$ of all stars present in one isochrone gives the total $N_{\rm Lyc}$ of the stellar population at a given time. Depending on the metallicity a fraction of this flux is absorbed by dust immediately (30\% is assumed to be absorbed for metallicities $\ge$ 0.008, no absorption is applied for lower-metallicity environments, following Mezger 1978, see also Weilbacher \etal 2000). From the remaining $N_{\rm Lyc}$ flux of an isochrone we calculate the gaseous continuum emission and the hydrogen line fluxes as described in Kr\"uger \etal (1995) and Weilbacher \etal (2000) assuming ionisation-recombination equilibrium (Osterbrock case B). The detailed formulae were already presented in Kr\"uger \etal (1995). For instance, the $N_{\rm Lyc}$ is converted into an $\rm H_\beta$ flux using \begin{center} $\rm F(H_\beta)=4.757 \times 10^{-13} \cdot ~N_{\rm Lyc}$. \end{center} Emission line fluxes for elements other than H are calculated from the line ratios relative to $\rm H_\beta$,  given in Table 1.

For low metallicity gas, line ratios for an exhaustive set of strong forbidden and allowed transitions from UV through NIR are obtained from the extensive observational database of Izotov \etal (1994, 1997) and Izotov \& Thuan (1998), subdivided into the metallicity bins covered by our SSP model grid, as given in the Z1- and Z2-columns in Table 1, for $Z~ {\rm =0.0004 ~and ~}Z{\rm =0.004}$ respectively.

Line ratios in reasonably metal-rich gas ($Z~{\rm =0.008, ~}Z_{\odot}, {\rm ~and ~}Z{\rm =0.05}$) are taken from Stasi\'nska (1984) as they have been shown to be in good agreement with Galactic HII region data (Sivan \etal 1986). No further distinction is made between the 3 metallicities $Z{\rm=0.008, ~}Z_{\odot}, {\rm ~and ~}Z{\rm =0.05}$, as galactic HII regions do show this full range of metallicities and their line ratios at fixed metallicity show considerable scatter.

The line ratios, of course, do vary with electron temperature, number density and/or the filling factor of the ionized gas. However, we aim to describe typical environments at the different metallicities. We have investigated the dispersion of line ratios within each of our metallicity bins $Z1$, $Z2$ and $Z3-Z5$ and the impact on integrated magnitudes in the broad band filters we consider. We find that, on average, the scatter in the line ratios is $\la$ 30\% (with few exceptions for weak lines). This scatter translates into a scatter in magnitude of at most 0.2 mag for solar metallicity, and up to 0.4 mag for the lowest metallicity. Compared with an error of 0.4 mag and 1.0 mag, respectively, by not taking gaseous emission into account at all, we consider this an improvement.  However, we would like again to emphasize, that our models are meant to improve upon models neglecting the gaseous emission in early evolutionary stages in terms of broad band luminosities and colors. For star clusters etc. that have detailed spectral information to determine all their ionising parameters, photoionisation codes will allow for a more precise individual description of their line strengths and their impact on colors.

\begin{table}[htbp]
\begin{center}
\caption{Non-hydrogen emission lines and their line strengths, normalized to ${\rm H_\beta}$ line strength, as a function of metallicity ($Z{\rm 1=0.0004}$, Z${\rm 2=0.004}$, Z${\rm 3=0.008}$, Z${\rm 4=0.02=}$ $Z_\odot$, $Z$${\rm 5=0.05}$).}
\begin{tabular}{|c|c|c|c|c|}
\hline
 & & & & \\ 
 Line & $\lambda$ [{\AA}] & $\frac{F_L}{F_{\rm H_\beta}}$ & $\frac{F_L}{F_{\rm H_\beta}}$ & $\frac{F_L}{F_{\rm H_\beta}}$\\
 & & & & \\
 & & $Z$1 & $Z$2 & $Z$3 -- $Z$5 \\
 & & & & \\ 
\hline
 & & & & \\
${\rm [CII]}$         & 1335.00&0.000&0.000&0.110\\
${\rm [OIII]}$        & 1663.00&0.000&0.058&0.010\\
${\rm [CIII]}$        & 1909.00&0.000&0.000&0.180\\
${\rm [NII]}$         & 2141.00&0.000&0.000&0.010\\
${\rm [CII]}$         & 2326.00&0.000&0.000&0.290\\
${\rm [MgII]}$        & 2798.00&0.000&0.310&0.070\\
${\rm [OII]}$         & 3727.00&0.489&1.791&3.010\\
${\rm [Ne III]}$      & 3869.00&0.295&0.416&0.300\\
${\rm H_\zeta ~+ ~[HeI]}$& 3889.00&0.203&0.192&0.107\\
${\rm H_\epsilon ~+ ~[NeIII]}$& 3970.00&0.270&0.283&0.159\\
${\rm [He I]}$        & 4026.00&0.015&0.015&0.015\\
${\rm [SII]}$         & 4068.60&0.005&0.017&0.029\\
${\rm [SII]}$         & 4076.35&0.002&0.007&0.011\\
${\rm [OIII]}$        & 4363.00&0.109&0.066&0.010\\
${\rm [HeI]}$         & 4471.00&0.036&0.036&0.050\\
${\rm [ArIV] ~+ ~[HeI]}$& 4711.00&0.010&0.014&0.000\\
${\rm [OIII]}$        & 4958.91&1.097&1.617&1.399\\
${\rm [OIII]}$        & 5006.84&3.159&4.752&4.081\\
${\rm [NI]}$	        & 5199.00&0.003&0.010&0.030\\
${\rm [NII]}$         & 5755.00&0.000&0.000&0.010	   \\
${\rm [HeI]}$         & 5876.00&0.096&0.108&0.140\\
${\rm [OI]}$	        & 6300.00&0.008&0.041&0.130\\
${\rm [SIII]}$        & 6312.00&0.009&0.017&0.030\\
${\rm [NII]}$         & 6548.05&0.005&0.059&0.136\\
${\rm [NII]}$         & 6583.45&0.015&0.175&0.404\\
${\rm [HeI]}$         & 6678.00&0.026&0.030&0.030\\
${\rm [SII]}$         & 6716.00&0.037&0.188&0.300\\
${\rm [SII]}$         & 6730.00&0.029&0.138&0.210\\
${\rm [HeI]}$         & 7065.00&0.028&0.023&0.040\\
${\rm [ArIII]}$       & 7135.79&0.027&0.071&0.035\\
${\rm [OII]}$         & 7319.99&0.012&0.027&0.026\\
${\rm [OII]}$         & 7330.73&0.007&0.014&0.014\\
${\rm [ArIII]}$       & 7751.11&0.067&0.176&0.086\\
${\rm [SIII]}$        & 9068.60&0.000&0.510&0.945     \\
${\rm [SIII]}$        & 9530.85&0.000&0.000&0.365	   \\
${\rm [SII]}$         &10286.73&0.000&0.000&0.048	   \\
${\rm [SII]}$         &10320.49&0.000&0.000&0.058\\
${\rm [SII]}$         &10336.41&0.000&0.000&0.054\\
 & & & & \\
\hline
\end{tabular}
\end{center}
\end{table}

In including the gaseous line emission our models go beyond Leitherer \etal 's {\sc starburst99} models which do include the gaseous continuum but no emission lines. Smith \etal (2002) will soon provide an updated version of the {\sc starburst99} models that also include emission lines with particular focus on WR star diagnostics. We recall that while {\sc starburst99} is based on stellar evolution input physics from the Geneva group, our models use Padova isochrones. 

\subsection{The role of white dwarfs} The inclusion of white dwarfs (WD)s changes the magnitudes of passbands redwards of $U$ by up to few times 0.001 mag. UV- and FUV-passbands show changes of generally 0.01 mag. However, even these values are likely to be upper limits -- at least in the description of star clusters -- since their derivation is based on the assumption of the presence of a sufficiently large amount of interstellar matter (ISM) to be ionized. 

In the case of WDs in star clusters, two aspects are of importance: First, WDs only appear after a delay of roughly 500 Myr, when SNe had more than enough time to blow away the interstellar material from a star cluster region. Second, the phase of planetary nebulae (PNe), when the region surrounding the WD is again refilled with gaseous material from the ejected envelopes until the ejected material is dispersed into the ISM, is fairly short in comparison with the time steps available ($\sim$ 25,000 yrs vs. 4,000,000 yrs). AS a result, the filling factor (and hence the average density) of the ISM during phases when WDs could contribute to the emission line flux is expected to be well below 1, resulting in a negligible contribution of gaseous emission due to WDs to broad band magnitudes. In addition, PNe in globular clusters seems to be surprisingly rare (Jacoby \etal 1997). 

In the case of galaxies, two scenarios might occur: Either the galaxy is gas-poor, hence not much matter is available that could be ionized, and the situation is similar to that of star clusters with low or negligible contributions of WDs. Or the galaxy is gas-rich, in which case the emission is probably dominated by star formation anyway, as described with our models presented here.

\subsection{Filter Systems}
In addition to the Standard Johnson $U~B~V~R~I~J~H~K~L$, Thuan \& Gunn $g~r$, Kron \& Koo $U^+~J^+~F^+~N^+$, Washington $C~M~T1~T2$, Str\"omgren $u~v~b~y$, {\sl HST} WFPC2 F160BW . . . F814W filters already included in Schulz \etal (2002), we here additionally include the {\sl HST} WFPC2 medium band filters F410M, F467M and F547M, the NICMOS filters F110W, F160W, F205W and F222M, and the ACS/WFC broad band filters F435W . . . F850LP. 

Calibrations for the ACS filters have been performed according to Gilliland 2002 ({\sl priv. comm.}). 

\section{Time Evolution of SSP Spectra with Gaseous Emission}
\begin{figure}[ht]
\includegraphics[angle=-90,width=\columnwidth]{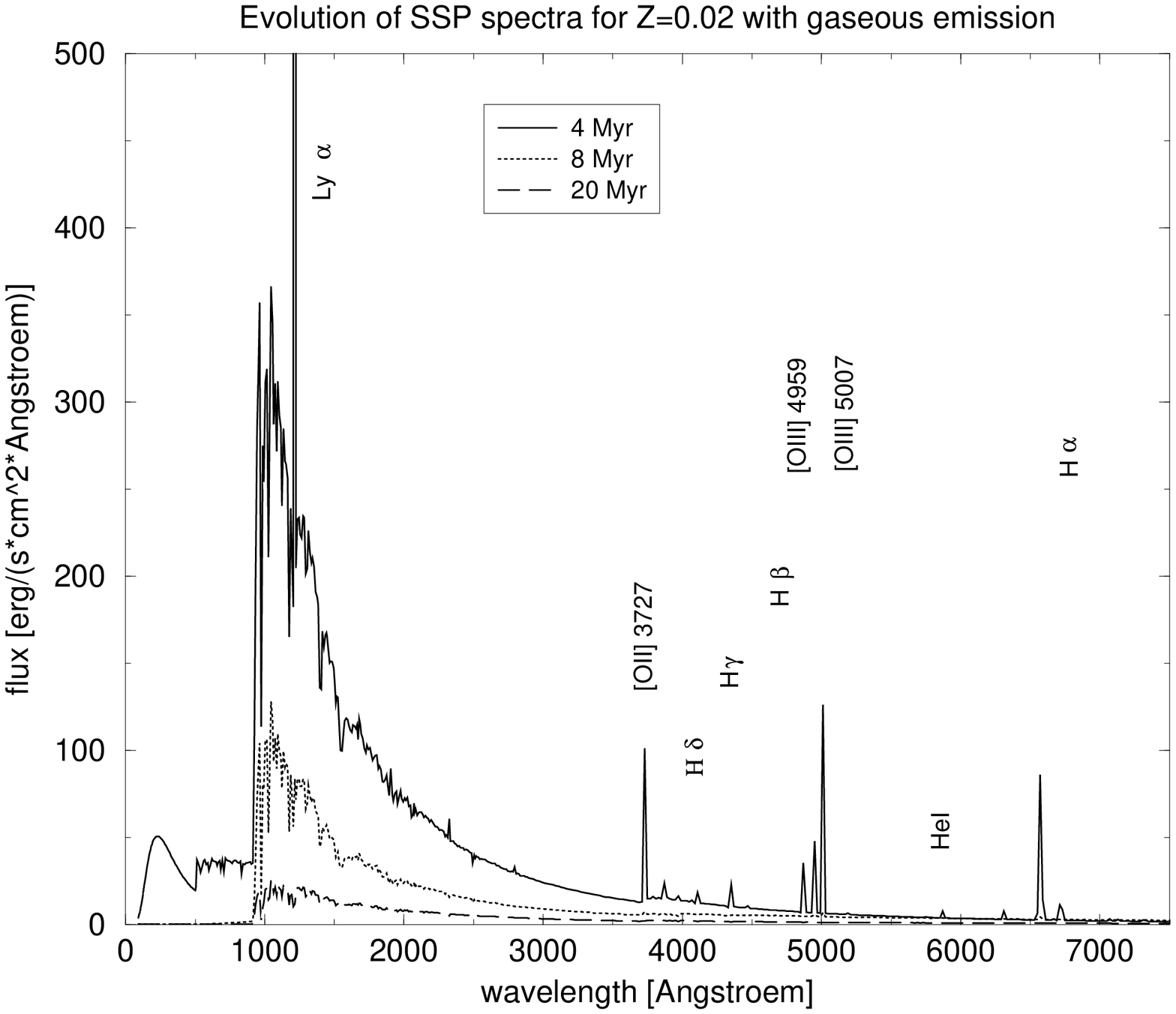}
\includegraphics[angle=-90,width=\columnwidth]{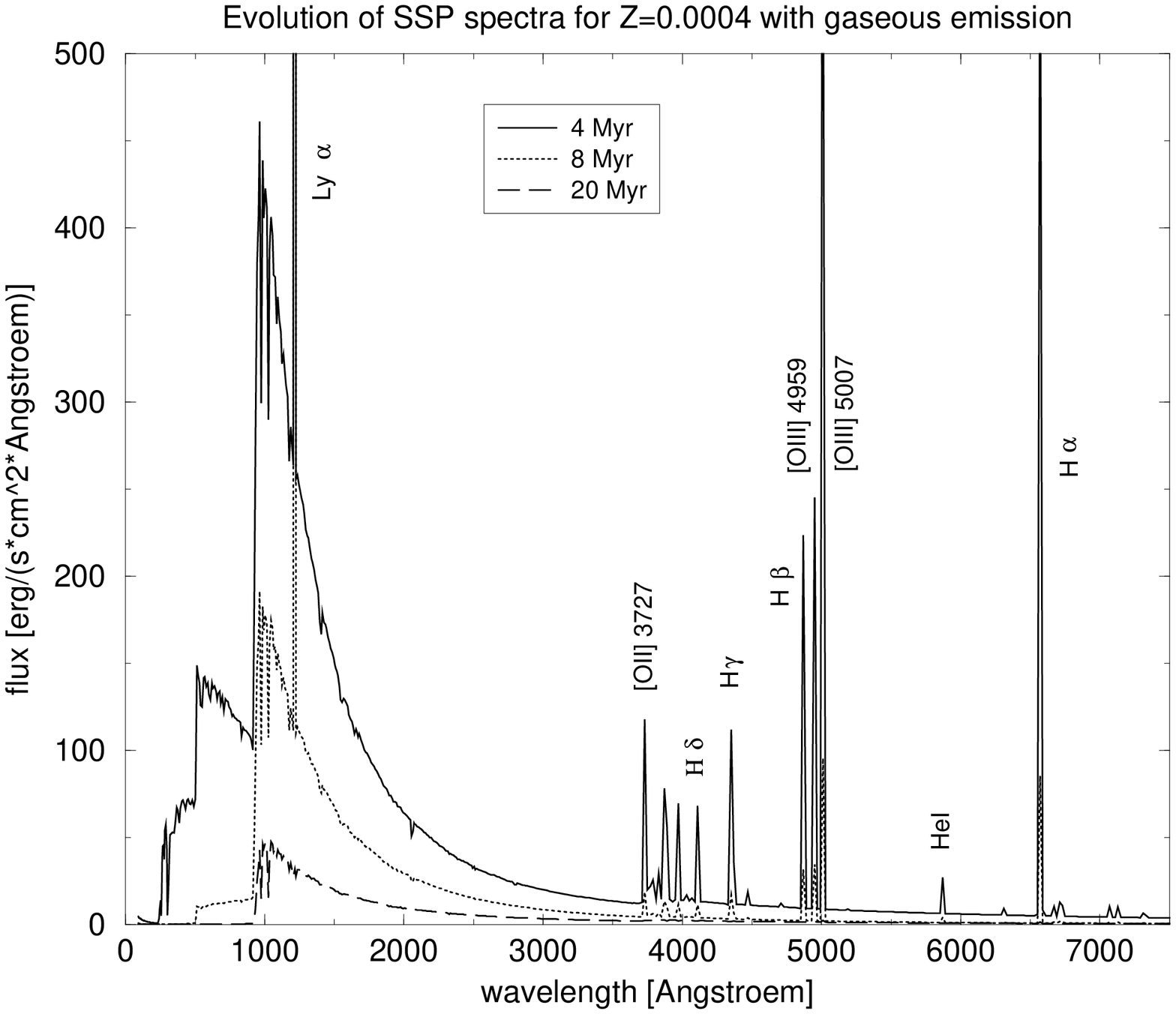}
\caption{Spectrum in terms of flux $F_{\lambda}$ as a function of wavelength $\lambda$ at 3 different times for an SSP of solar metallicity {\bf (a)} and metallicity $Z~ {\rm =0.0004}$ {\bf (b)}, both with Salpeter IMF.}
\label{fig_spec}
\end{figure}

\begin{figure}[ht]
\includegraphics[angle=-90,width=\columnwidth]{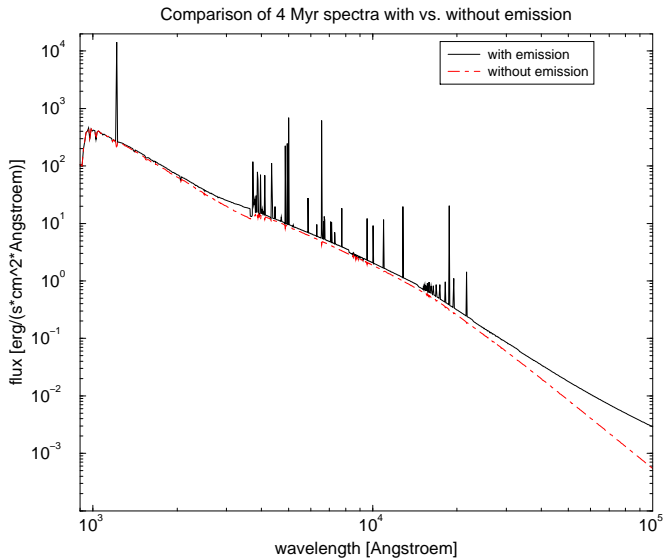}
\caption{Spectrum in terms of flux $F_{\lambda}$ as a function of wavelength $\lambda$ for an SSP with $Z=0.0004$ and Salpeter IMF at an age of 4 Myr, with and without gaseous emission, in log-log presentation for better visibility.}
\label{fig_spec2}
\end{figure}

The new SSP models presented here extend to younger ages and have a better time resolution as compared to Schulz \etal (2002). Time resolution now is 4 Myr up to an age of 2.35 Gyr, and 20 Myr for larger ages. The youngest age available now is 4 Myr as compared to 140 Myr in Schulz \etal. In Fig. \ref{fig_spec} we present the early evolution of the UV -- optical parts of SSP spectra of solar and low metallicity, respectively, for comparison. The much stronger emission lines in the low metallicity spectra are clearly seen as well as some differences in line ratios, as e.g. that of [OII]3727 relative to the adjacent Balmer lines ${\rm H_{\delta}}$ and ${\rm H_{\gamma}}$. The time evolution and metallicity dependence of the line strengths will further be discussed in Section \ref{sec_impact_bb}. Fig. \ref{fig_spec2} shows spectra of a young low metallicity ($Z=0.0004$) SSP over a longer wavelength range.


\section{Impact of Gaseous Emission on Broad Band Luminosities and Colors}
\label{sec_impact_bb}
\subsection{Relative contributions of the gaseous emission -- lines and continuum -- to broad band fluxes}

\begin{figure}[ht]
\includegraphics[angle=-90,width=\columnwidth]{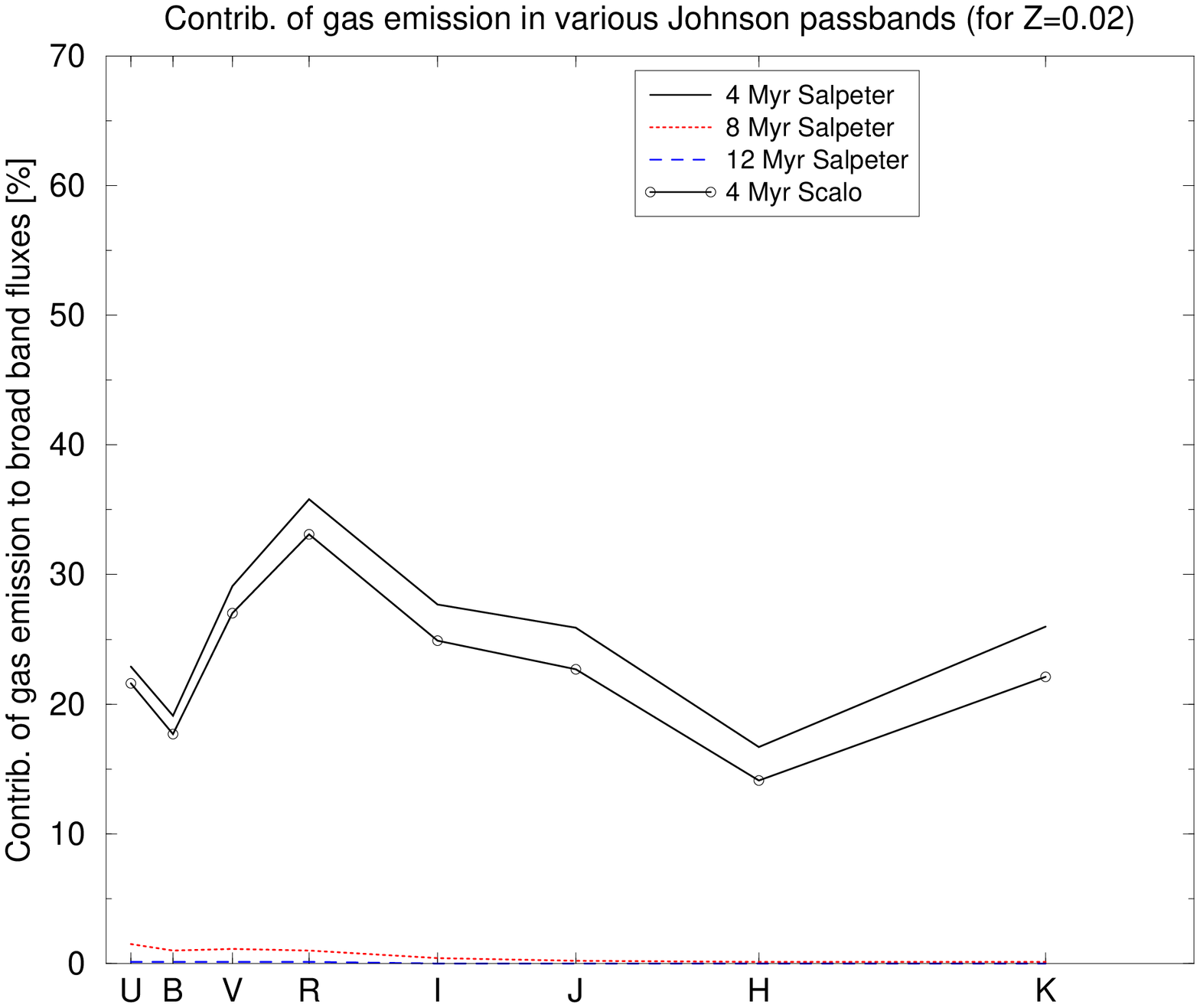}
\includegraphics[angle=-90,width=\columnwidth]{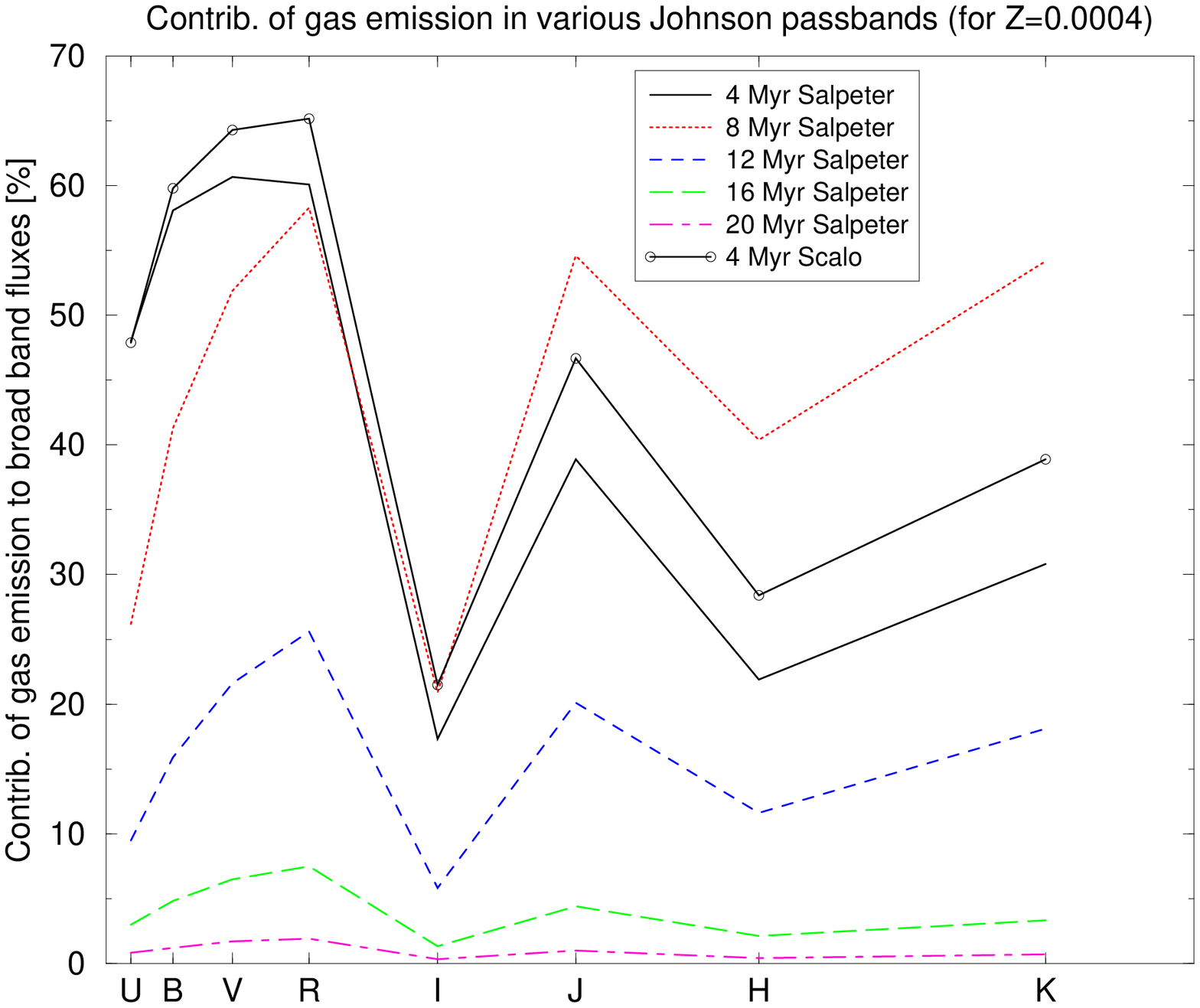}
\caption{Relative contributions of the gaseous emission -- lines and continuum -- to broad band fluxes $U$ . . .  $K$ at solar metallicity {\bf (a)} and low metallicity $Z~ {\rm = 0.0004}$ {\bf (b)}.}.
\label{fig_wl_con}
\end{figure}

\begin{figure}[ht]
\includegraphics[angle=-90,width=\columnwidth]{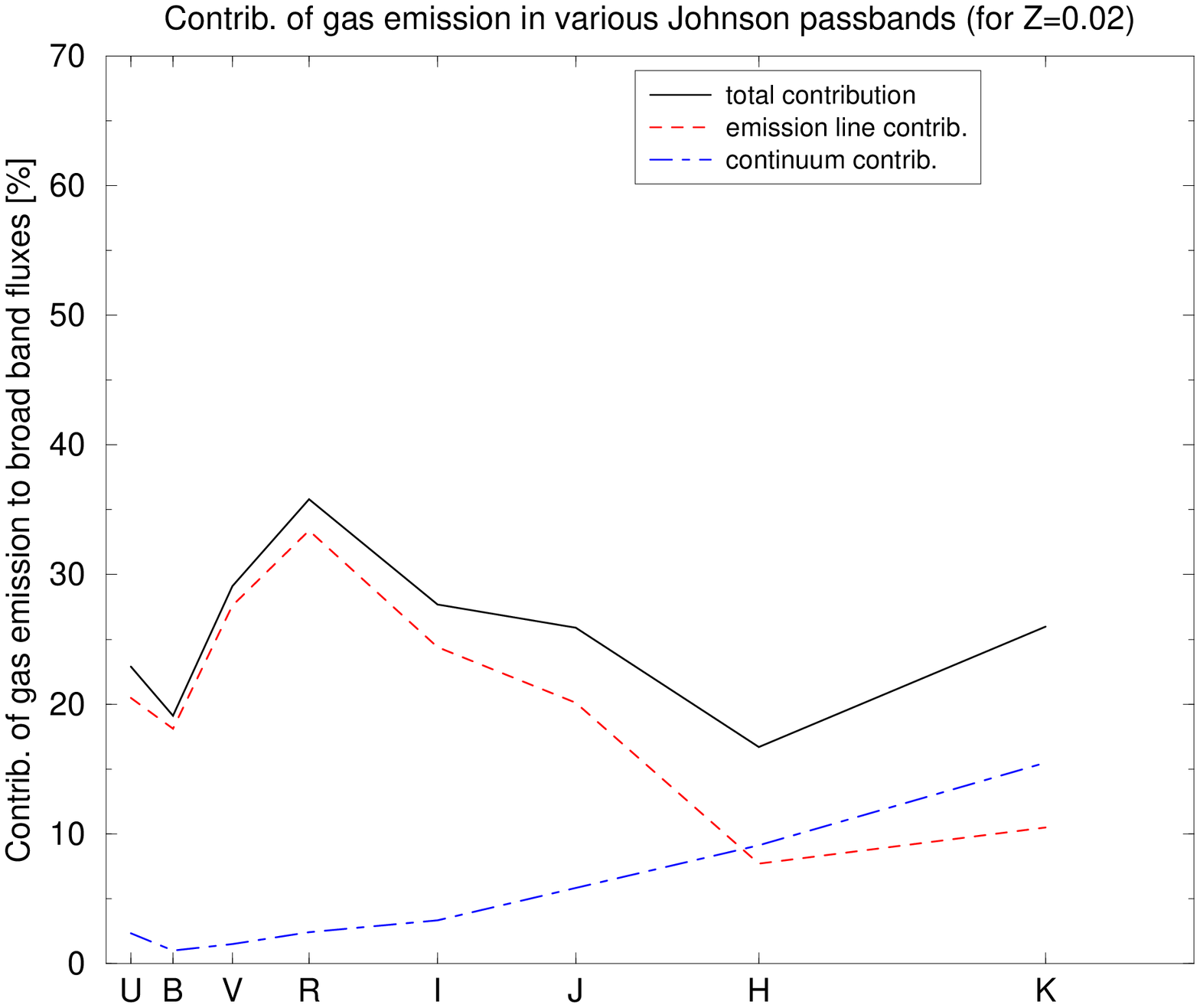}
\includegraphics[angle=-90,width=\columnwidth]{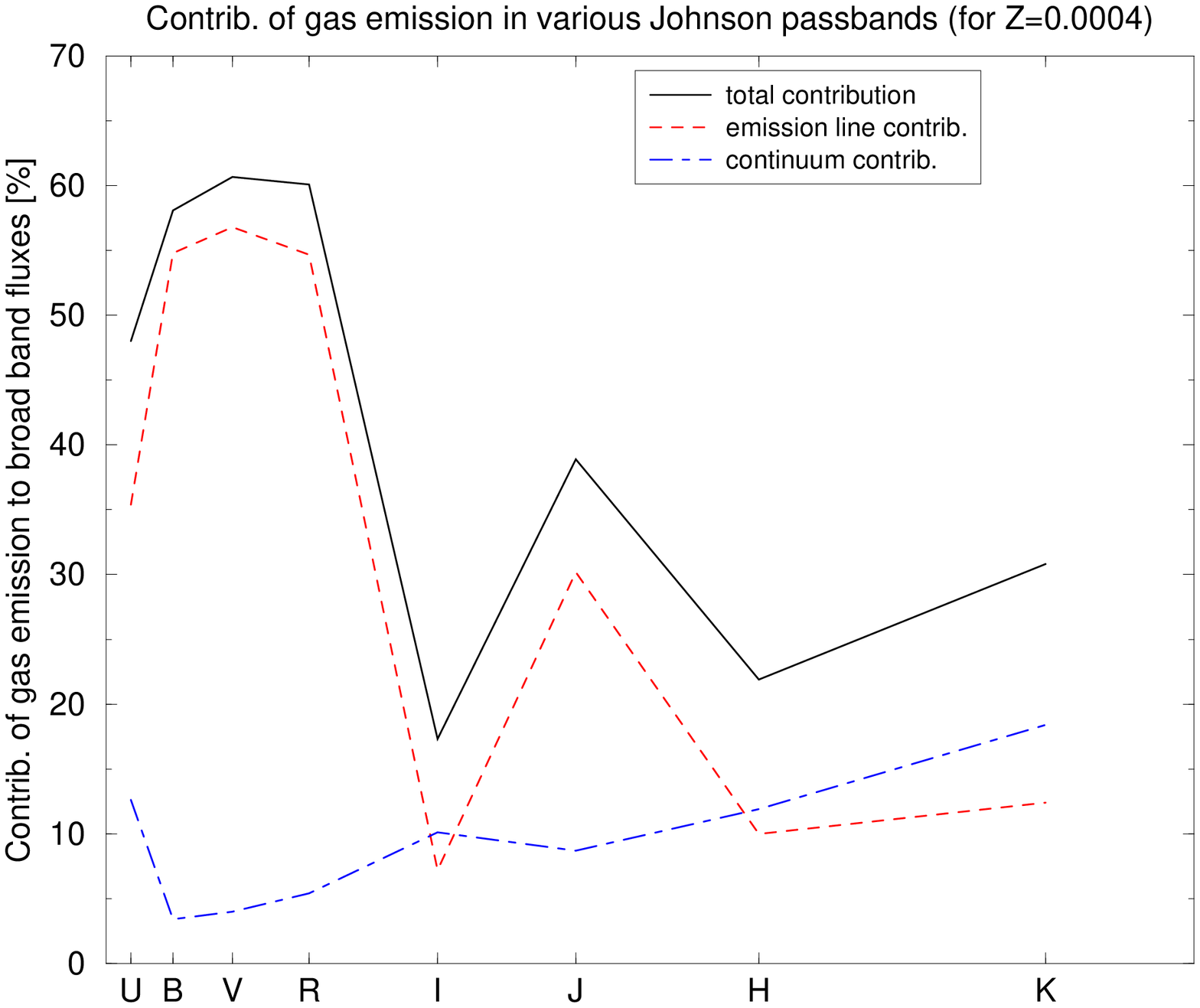}
\caption{Decomposition of the total gaseous emission in terms of lines and continuum contributions in various filters bands $U$ . . .  $K$ at an age of 4 Myr at solar metallicity {\bf (a)} and low metallicity $Z~ {\rm = 0.0004}$ {\bf (b)}.}
\label{fig_con_decomp}
\end{figure}

In Fig. \ref{fig_wl_con}, the relative contributions of the gaseous emission to broad band fluxes in the Johnson passbands $U$ . . . $K$ are shown for some young ages, again for solar and low metallicity. The metallicity dependence visible in the $B$-band is due to strong oxygen and neon lines. ${\rm H_\beta}$, ${\rm H_\gamma}$ and ${\rm H_\delta}$ contribute significantly as well. The $V$-band contribution is dominated by strong oxygen lines (and to a lesser extent helium, nitrogen and sulphur), the strengths of which are metallicity dependent. The strong contribution in the $R$-band is caused by ${\rm H_\alpha}$, which is located close to the throughput maximum of the Johnson $R$ filter. The $I$-band is dominated by sulphur lines for high metallicities (predominantly the ${\rm [SIII]}\lambda9068.60$ and ${\rm [SIII]}9530.85$ lines, both being moderately strong and close to the I-band throughput maximum), which are absent in the case of $Z~ {\rm =0.0004}$. The remaining lines are weak or far from the maximum of the filter response function, resulting in a prominent dip in Figures \ref{fig_wl_con}b/\ref{fig_con_decomp}b. Only lines from the Paschen- and Brackett-series with small relative flux contributions are present in the NIR bands $J$, $H$ and $K$. Here, the contribution from the continuum emission becomes important.

To interpret the differences between Scalo and Salpeter IMF, as seen in Fig. \ref{fig_wl_con}, two points are of importance: the contributions of the stars to the stellar continuum and the contributions to the ionising photons. The stellar continuum is dominated by stars around 9 $M_\odot$, more or less independent of IMF and metallicity. However, the stars responsible for the ionising photons span a larger mass range in the low metallicity case as compared to solar metallicity, providing more ionising photons at low metallicity. In combination with the lower overall stellar luminosity for the Scalo IMF in comparison to the Salpeter IMF, this leads to an enhanced relative emission contribution for the low metallicity Scalo IMF, and a reduced contribution for the solar metallicity case, both compared to the respective Salpeter values.

While Fig. \ref{fig_wl_con} shows the overall emission contribution of the gas to broad band fluxes $U$ . . . $K$, this is decomposed into its line and continuum components in Fig. \ref{fig_con_decomp}. This decomposition of the relative flux contribution of the gas to the fluxes in various filters shows that -- for all metallicities -- lines dominate in optical filters while continuum emission comes into play in the NIR bands. A similar result was already reported by Kr\"uger \etal (1995) in the context of evolutionary synthesis models for BCD galaxies. Like the hydrogen lines, the continuum emission is not directly metallicity-dependent, but indirectly via the higher ionising flux of lower metallicity SSPs and the dust absorption rate $f(Z)$ for ionising photons.

\subsection{Time evolution of gaseous emission contributions}

\begin{figure}[ht]
\includegraphics[angle=-90,width=\columnwidth]{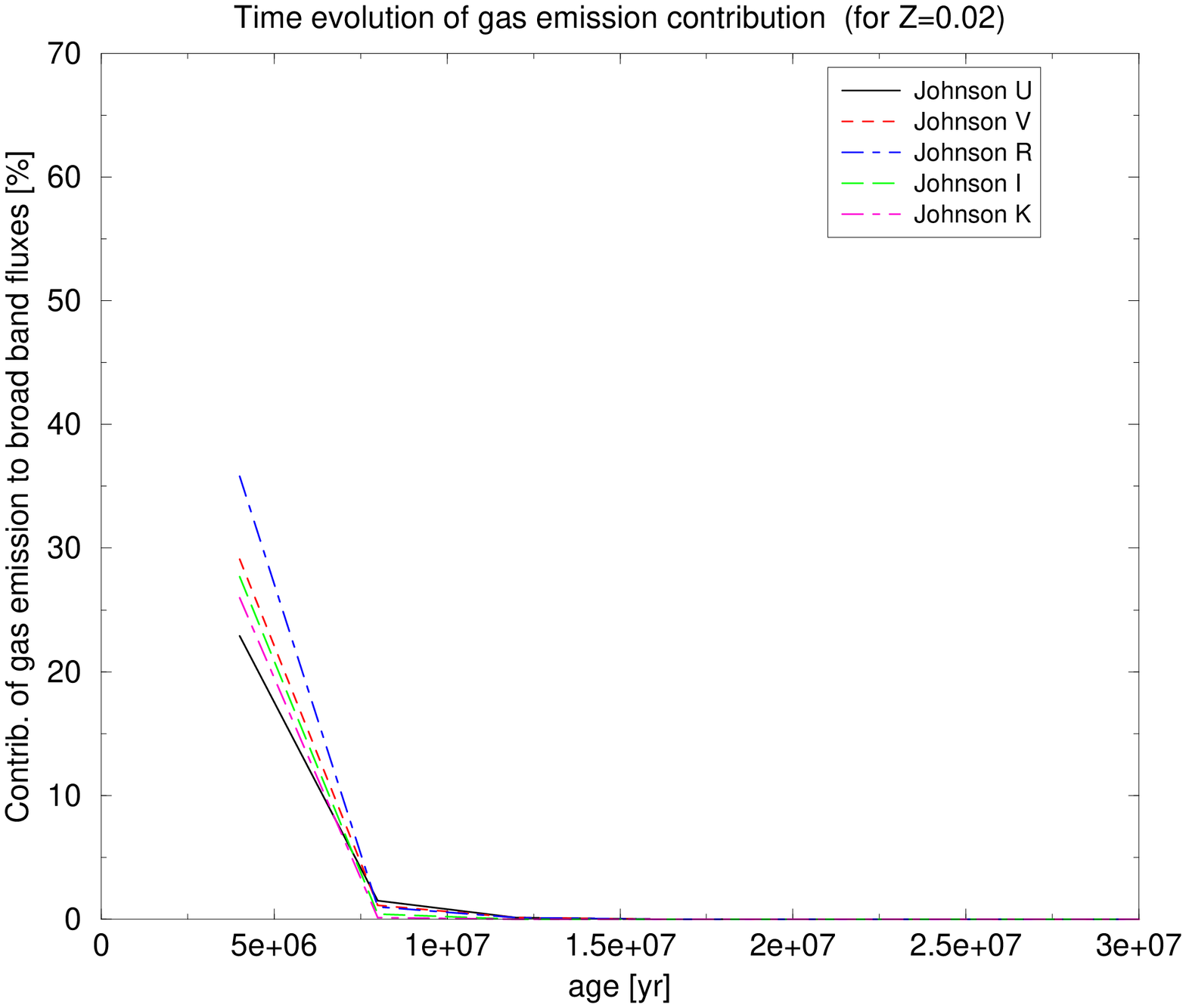}
\includegraphics[angle=-90,width=\columnwidth]{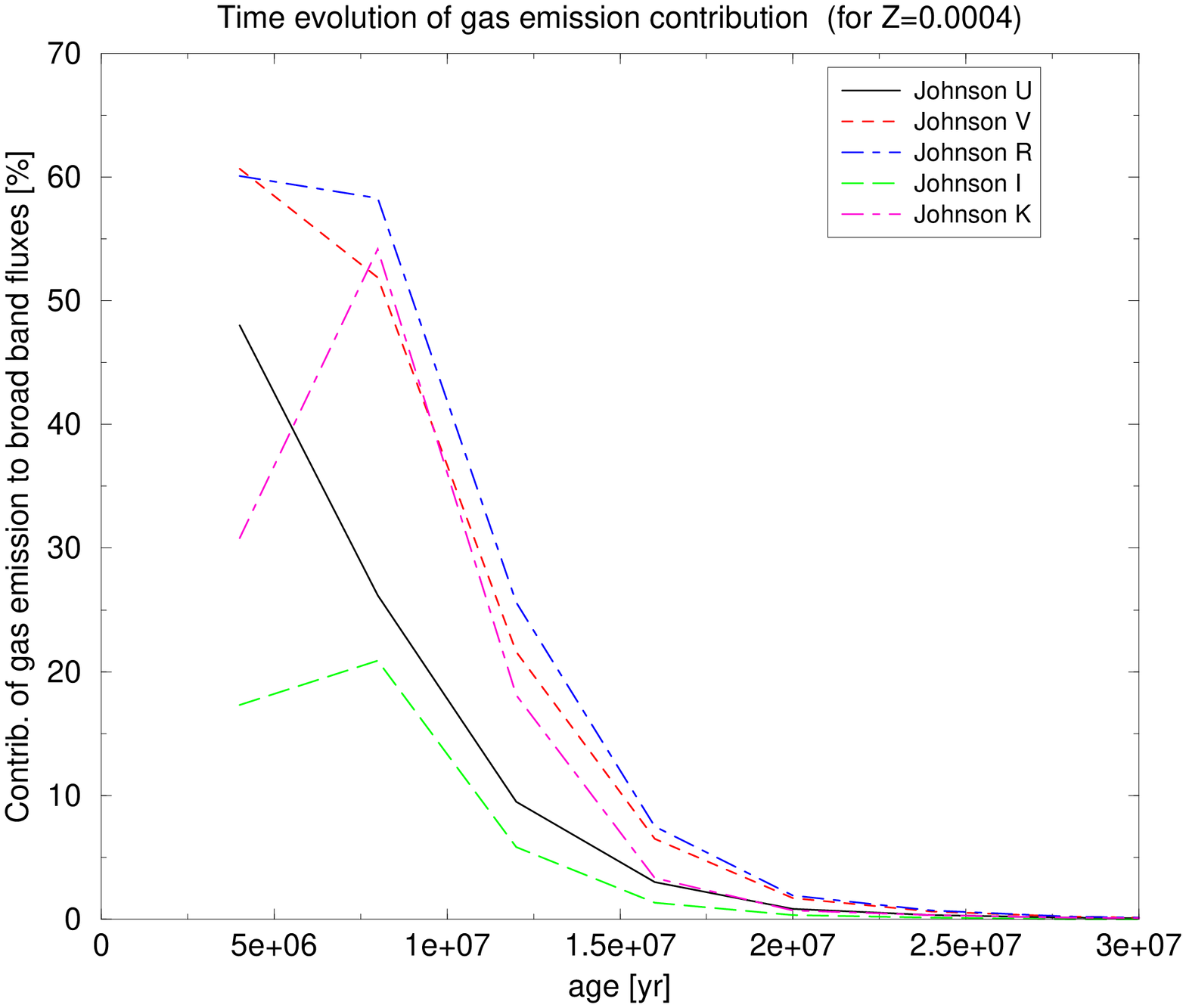}
\caption{Time evolution of the gaseous emission contribution to broad band fluxes $U$, $B$, $V$, $I$, and $K$ at solar metallicity {\bf (a)} and low metallicity $Z~ {\rm = 0.0004}$ {\bf (b)}.}
\label{fig_time_con}
\end{figure}

As can be seen from Fig. \ref{fig_time_con}, the gaseous emission contribution is much stronger as well as longer-lasting at lower metallicity. In the Padova (as well as in Geneva) isochrones, the most massive stars have longer lifetimes at lower metallicities, as compared with solar. As discussed in Sect. \ref{sec_input_gas}, the population of massive stars gets hotter and more luminous, on average, at lower metallicities as compared to solar resulting in a higher output rate of ionising photons. In addition, the fraction of ionising photons directly absorbed by dust gets lower. Hence the relative contribution of gaseous emission e.g. to the $R$-band (dominated by ${\rm H_\alpha}$) starts from 35 \% at $Z_{\odot}$ and from 60 \% at $Z~ {\rm =0.0004}$ and decreases by a factor 3 within 5 and 13 Myr, respectively. At solar metallicity, the overall relative flux contribution of the ionised gas falls below 5 \%  in all bands at ages $\le$ 8 Myr. At our lowest metallicity, $Z~ {\rm =0.0004}$, it falls below 5 \% at 13 Myr ($I$-band), 15 Myr ($U$, $K$), and 18 Myr ($V$, $R$).

\subsection{Effects of gaseous emission on the color evolution of SSPs}

\begin{figure}[ht]
\includegraphics[angle=-90,width=\columnwidth]{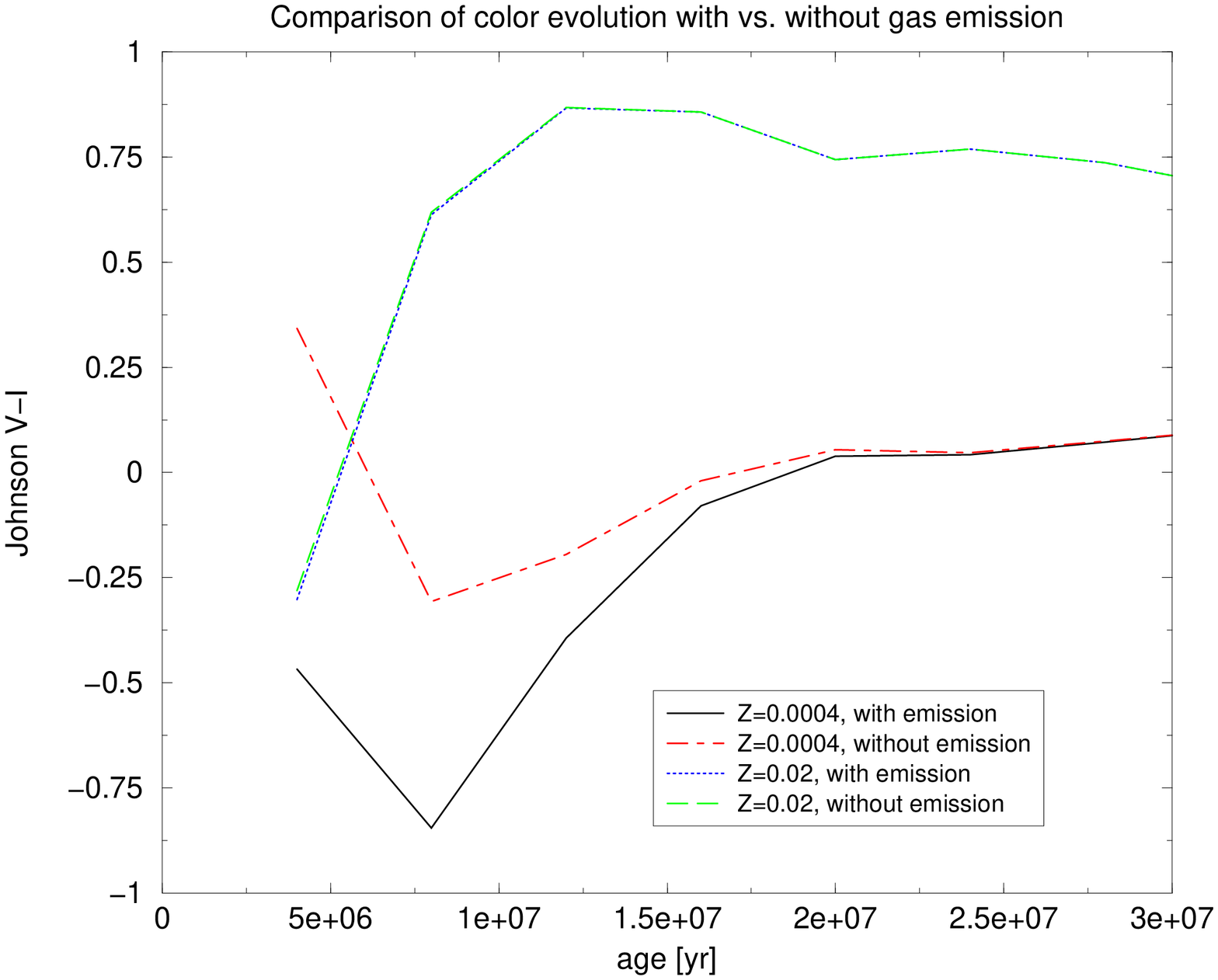}
\caption{Time evolution of the Johnson colors $V-I$ with and without gaseous emission at solar metallicity $Z~ {\rm = 0.02}$ and low metallicity $Z~ {\rm = 0.0004}$.}
\label{fig_VI}
\end{figure}
\begin{figure}[ht]
\includegraphics[angle=-90,width=\columnwidth]{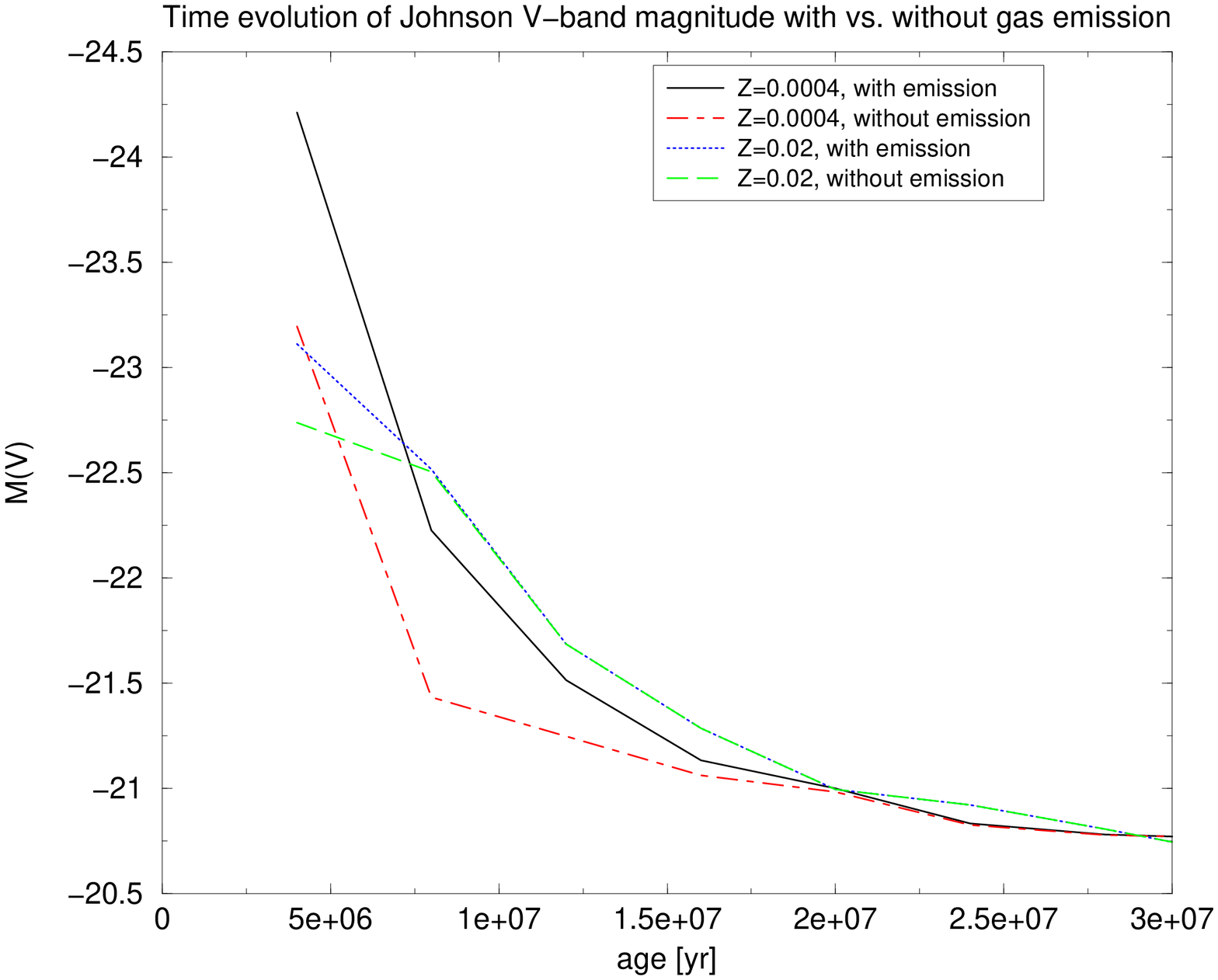}
\caption{Time evolution of the magnitude in the Johnson $V$-band with and without gaseous emission at solar metallicity $Z~ {\rm = 0.02}$ and low metallicity $Z~ {\rm = 0.0004}$ for SSPs of the same mass.}
\label{fig_MV}
\end{figure}

As seen in Fig. \ref{fig_VI} on the example of $V-I$ the inclusion of gaseous emission significantly affects the color evolution of SSPs at early stages, depending on the metallicity, and the luminosity of a cluster of given mass, as can be seen in Fig. \ref{fig_MV}. This also, of course, affects the age and mass determinations from observed colors and luminosities.

From an observed $V-I=-0.2$ of a very young cluster, e.g., ages of 4 and 7.5-12 Myr would be derived in case of solar and metallicity $Z~ {\rm =0.004}$, respectively, if the effects of gaseous emission were not taken into account. Including them appropriately, the same observed color $V-I=-0.2$ would result in ages 4 and 14.5 Myr, for the two metallicities. This shows that ages estimated on the basis of $V-I$, without taking gaseous emission into account, for low metallicity clusters younger than $\sim$ 15 Myr are underestimated by up to a factor $\sim$ 2. Differences decrease with increasing metallicity, and if gas emission is neglected at solar metallicity it has become negligible. In a similar way, masses estimated from the observed cluster $M_V$ with model mass-to-light ratios that do not account for gaseous emission contributions can be overestimated by a factor 2 at low metallicity.

Observed colors $V-I$ in the range [-0.35,-0.9] cannot at all be accounted for by models neglecting gaseous emission contributions.

\section{Discussion}
The aim of our models as presented here is to provide the spectral and photometric evolution of SSPs of various metallicities that also allow to describe the earliest stages of evolution when the ionising flux of a star cluster, a stellar (sub-)population etc. is strong enough to cause the surrounding gas to emit line and continuum radiation. Our motivation was that many {\sl HST} observations of young star cluster systems in interacting and merging galaxies indeed do show such very young clusters with broad band colors that cannot be interpreted without accounting for their gaseous emission. Spectroscopy of some of those bright clusters has explicitly revealed their emission lines, e.g. Schweizer \& Seitzer (1998) in NGC 7252, Colina \etal (2002) in NGC 4303, Maoz \etal (2001) in NGC 1512 and NGC 5248, ...

We used most recent compilations of stellar output rates in ionising photons. Note that, while finding considerable differences in the output rates of He-ionising photons, the latest and most sophisticated calculations by Smith \etal (2002) confirm the rates from Schaerer \& de Koter (1997) that we use for H-ionising photons.

Our focus is on the impact of gaseous emission on the early luminosity and color evolution. For low metallicities ($Z~ {\rm \le 0.004}$), we chose to use line ratios for the different metallicities of our SSP models as observed in HII regions of the respective metallicities. The line ratios for higher metallicities were taken from the theoretical models presented by Stasi\'nska (1984). Hence, our models should not be used for any analyses of line ratios. For this purpose we recommend the use of models that couple to photoionization codes like those of Charlot \& Longhetti (2001) or Moy \etal (2001).

In Schulz \etal (2002) we gave evolutionary and cosmological corrections for our SSP models without gas for convenient use in cosmological structure formation models or any kind of models that build up galaxies with extended star formation (and possibly also chemical enrichment) histories. As the changes caused by the inclusion of gaseous emission are constrained to the earliest evolutionary stages ($t \la$ 15 Myr) their effects on the cosmological and evolutionary corrections are only visible at redshifts extremely close to the redshift of the very onset of star formation. We therefore do not present cosmological and evolutionary corrections here and advise potential users to stay with the tables in Schulz \etal (2002).

\begin{figure}[ht]
\includegraphics[angle=-90,width=\columnwidth]{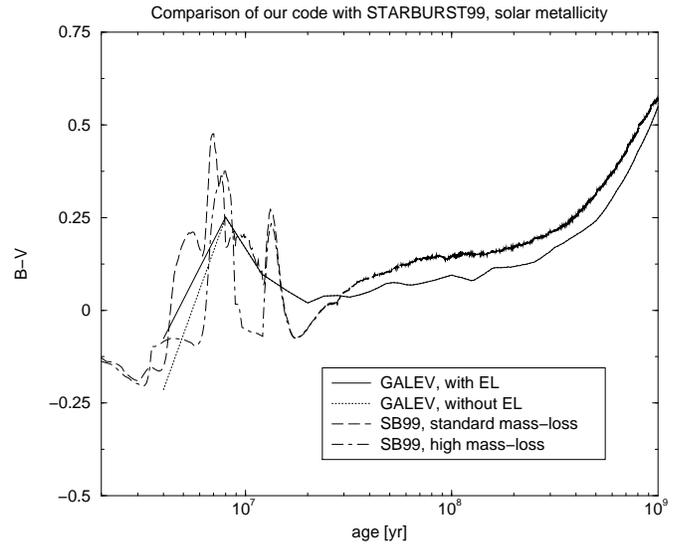}
\caption{Comparison of our models with {\sc starburst99} models with different mass-loss rates, for solar metallicity and the Johnson color $B-V$.}
\label{fig_SB}
\end{figure}

Comparison of our {\sc galev} model using Padova stellar input physics with results from Leitherer \etal's (1999) {\sc starburst99} code, using Geneva stellar tracks, during the first Gyr of evolution covered by the latter shows fairly good agreement. As an example we show in Fig. \ref{fig_SB} the time evolution of $B-V$. The differences are minor ($<$ 0.1 mag), and mainly due to the better time resolution of the {\sc starburst99} code. Comparison of colors including longer wavelengths, e.g. $V-I$ or $V-K$, show larger deviations that are due to different input physics, especially a different treatment of red supergiants and the omission of the TP-AGB phase in {\sc starburst99}.

\section{Structure and Description of Electronic Data-files}
Our model results are accessible at CDS and at http://www.uni-sw.gwdg.de/$\sim$galev/panders/ . The files have the following naming conventions:\\
\begin{itemize}
\item Each directory is labelled ssp\_model\_z{\sl{x\_y}}. Here {\sl{zx}} (x = 1 . . . 5) indicates the metallicity (z1 $\Leftrightarrow$ $Z~ {\rm =0.0004}$, z2 $\Leftrightarrow$ $Z~ {\rm =0.004}$, z3 $\Leftrightarrow$ $Z~ {\rm =0.008}$, z4 $\Leftrightarrow$ $Z~ {\rm =0.02}$, z5 $\Leftrightarrow$ $Z~ {\rm =0.05}$), and {\sl{y}} the assumed IMF $\Phi(m) \sim m^{\alpha}$ ('salpeter' $\Leftrightarrow$ ${\rm \alpha=-2.35}$ for all masses; 'scalo' $\Leftrightarrow$ ${\rm \alpha=-1.25}$ for $m \le 1 M_{\odot}$, ${\rm \alpha=-2.35}$ for $1 M_{\odot} < m \le 2 M_{\odot}$ and ${\rm \alpha=-3.00}$ for $2 M_{\odot} < m$).
\item Each directory contains a file named 'plotsgal.dat' containing the cluster mass as a function of age, a file named 'galspec.dat' containing the integrated cluster spectra as a function of age, and a file named 'magnitudes.dat' containing the integrated magnitudes of a cluster as a function of age in various passbands.
\item Further information is included in the README file.
\end{itemize}

\section{Conclusions and Outlook}
We present an update of the evolutionary spectral synthesis models for SSPs of various metallicities presented in Schulz \etal (2002). Using the same stellar isochrones provided by the Padova group, including the TP-AGB phase important for $V-I$ and $V-K$ colors of SSPs in the age range $10^8$ . . . $10^9$ yr, we add the gaseous emission contributions, both lines and continuum. They are very important not only for the spectral evolution but also for the photometric evolution in terms of broad band colors in early evolutionary stages, as long as ionising stars are alive. We use the best available output rates for H-ionising photons and observationally supported emission line ratios appropriate for the different metallicities. This allows us to extend our SSP models towards younger ages than accessible before. We provide an extensive set of electronic data for the time evolution of UV through NIR spectra of SSPs of metallicities $Z~ {\rm =0.0004,~0.004,~0.008,~0.02=}Z_{\odot}, {\rm ~and ~0.05}$, for Salpeter and Scalo IMFs, from ages as young as 4 Myr all through 14 Gyr in timesteps of 4 Myr until an age of 2.35 Gyr and of 20 Myr thereafter. We also provide the luminosity and color evolutions in a large number of ground- and space-based filter systems including Johnson, Washington, {\sl HST} WFPC2, NICMOS and ACS/WFC.

Our results also allow for easy superposition of SSPs of various ages -- and possibly metallicities -- to describe galaxies with all possible star formation and metal enrichment histories. They also are readily included into dynamical and cosmological galaxy or structure formation scenarios that contain a star formation criterium.

The evolutionary and cosmological corrections we gave in Schulz \etal (2002) are not affected by gaseous emission as it is constrained to stages earlier than 15 Myr, i.e. to redshifts extremely close to the redshift at which star formation sets in.

\acknowledgement
PA is partially supported by DFG grant Fr 916/11-1. We like to thank the anonymous referee for his or her suggestions that helped to improve the presentation of our results.

\end{document}